# Morphisms of Coloured Petri Nets

**Joachim Wehler**

Ludwig-Maximilians-Universität München, Deutschland
joachim.wehler@gmx.net

**Abstract.** We introduce the concept of a morphism between coloured nets. Our definition generalizes Petris definition for ordinary nets. A morphism of coloured nets maps the topological space of the underlying undirected net as well as the kernel and cokernel of the incidence map. The kernel are flows along the transition-bordered fibres of the morphism, the cokernel are classes of markings of the place-bordered fibres. The attachment of bindings, colours, flows and marking classes to a subnet is formalized by using concepts from sheaf theory. A coloured net is a sheaf-cosheaf pair over a Petri space and a morphism between coloured nets is a morphism between such pairs. Coloured nets and their morphisms form a category. We prove the existence of a product in the subcategory of sort-respecting morphisms. After introducing markings our concepts generalize to coloured Petri nets.

## 1   Introduction

The concept of a morphism between ordinary nets is well-defined: A morphism maps the nodes and respects adjacency and orientation. The present paper generalizes this definition to coloured nets.

For a morphism between ordinary nets the inverse image of a place (resp. transition) is a place-bordered (resp. transition-bordered) subnet. This property has a simple translation into the language of topology. The bipartite structure of an undirected net introduces two Petri topologies on the set of nodes. One of them, the P-topology, has open sets the place-bordered subnets. The other has open sets the transition-bordered subnets. Each Petri topology expresses all graph theoretical properties of an undirected net. In particular, a map respects the adjacency, iff it is continous with respect to one – and hence to both – Petri topologies. Chapter 3 *The two Petri topologies of a net* surveys some of the topological properties of Petri spaces and their continous maps.

While adjacency is a topological property, the orientation of a net supersedes topology. Besides their orientation, arcs of a coloured net have weights. An arc weight maps binding-elements to token-elements. These additional properties of a coloured net express a second, algebraic structure. It consists of monoids and modules and their corresponding morphisms. These algebraic objects are attached to the open resp. closed subsets of the underlying undirected net. They arise from transitions, being closed singletons, and places, being open singletons. *Sheaf* and *cosheaf* are suitable mathematical concepts, to endow a topological space with a family of algebraic objects.

Chapter 4 *Coloured nets as sheaf-cosheaf pair* introduces these concepts. It formalizes a coloured net as a pair of a sheaf and a cosheaf together with two morphism between them (Definition 4.9). From the kernel of the incidence morphism derives the sheaf of flows. It attaches to every closed subset of the Petri space the flows of this transition-bordered subnet. Analogously, from the cokernel of the incidence morphism derives the cosheaf of marking classes. Two markings, which are potentially reachable from each other, are considered equivalent. The cosheaf of marking classes attaches to every open subset the marking classes of this place-bordered subnet. One essential property of a flow is embodied into the sheaf definition: A flow of a net restricts to a flow on every transition-bordered subnet. Similarly, a marking class of a place-bordered subnet extends to a marking class on every embedding net. This property is embodied into the cosheaf definition. Chapter 4 requires the willingness of the reader to consider some concepts from mathematics, which are new in computer science.

Chapter 5 *Morphisms of coloured nets* defines such morphisms as a morphism between the two sheaf-cosheaf pairs (Definition 5.1). The definition is inspired by the work of Lakos. The morphism has three components: A continous map between the underlying Petri spaces, a map between the two sheaves of flows and a map between the two cosheaves of marking classes. All three maps must be compatible with respect to the incidence morphisms. A morphism between coloured nets is not necessarily sort-respecting. Special types of morphisms are embeddings and abstractions. After adding initial markings to the coloured nets we define morphisms between two Petri nets as morphisms of the underlying coloured nets, which map the initial marking as well as certain activated occurrence sequences (Definition 5.8). We prove that a morphism with open image always extends to a morphism of Petri nets (Proposition 5.7). Morphisms in the sense of Winskel can be represented - after a modification of their image - by our morphisms (Proposition 5.13). This modification keeps the behaviour.

For morphisms of ordinary nets no categorical product is known. We prove in Chapter 6 *Products of coloured* nets the existence of a product in the category of coloured nets with rational coefficients and sort-respecting morphisms (Proposition 6.4). The product serves as the base for the categorical product of coloured Petri nets and sort-respecting morphisms (Proposition 6.6). Applying the diagonal construction we derive the existence of fibre products (Proposition 6.11).

All nets in this paper are assumed to be finite - concerning the number of transitions and places as well as concerning the number of their colour-elements. Nevertheless, the joint methods of sheaf theory and the theory of topological vector spaces allow to deal with infinite nets, too. This will be detailed elsewhere

## 2   Running example

We will illustrate the new concepts of the paper at the example from Figure 1. It shows a morphism between two nets. The source $N_X$ is an ordinary net, the target $N_Y$ is a coloured net. The coloured net has a single transition "a" with two binding-elements "$b_1$", "$b_2$" and a single place "u" with one token-element "c". Each binding-element consumes and creates two token-elements.

The topological part of the morphism is a continous map $f : X \longrightarrow Y$ between the underlying Petri spaces, which maps the nodes of $N_X$ onto the nodes of $N_Y$. We want to direct the attention to the fibre structure of $N_X$, which results from f. Hence we distinguish between T-fibres as inverse images of transitions and P-fibres over places. The set $\{ p_1, p_2, t_1, t_2, t_3, t_4 \}$ spans $N_{X,a} := f^{-1}(a) \subset N_X$, a transition-bordered T-fibre. The set $\{ p_3, p_4, t_5, t_6 \}$ spans the place-bordered P-fibre $N_{X,u} := f^{-1}(u) \subset N_X$. The map f refines a transition (resp. a place) of $N_Y$ by a transition-bordered (resp. place-bordered) subnet of $N_X$.

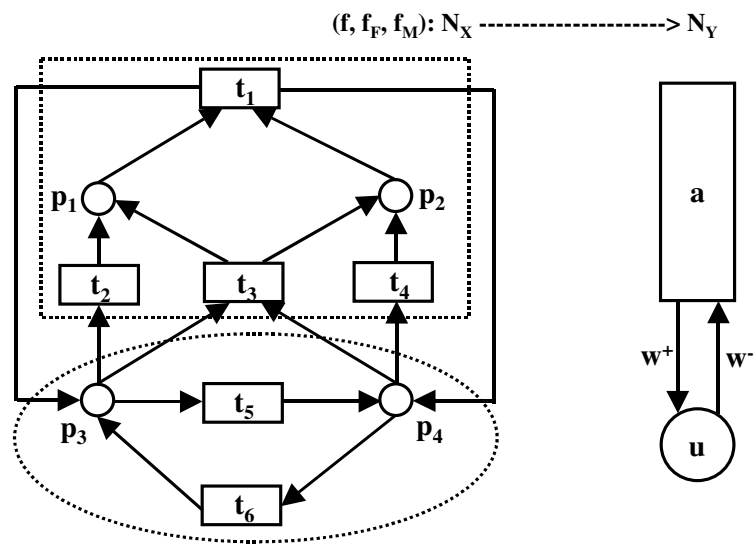

*Figure 1 Morphism onto a coloured net*

The transition-bordered fibre $N_{X,a}$ has two flows $\tau_1 := t_1 + t_3$ and $\tau_2 := t_1 + t_2 + t_4$. Following the proposal of Lakos ([Lak1997]) we consider them as the binding-elements of the whole T-fibre. We consider a marking as a linear functional on the places. Then the place-bordered fibre $N_{X,u}$ has one marking class $\pi = \left[ p_3^* \right] = \left[ p_4^* \right]$ with respect to potential reachability. We attach it to the whole P-fibre as its token-element. In order to extend the continous map f on the level of undirected nets to a morphism of coloured nets (f, $f_F$, $f_M$): $N_X \to N_Y$, we will introduce to other maps: The map $f_F$ will map bindings-elements of T-fibres to binding-elements of the target, and the map $f_M$ will map token-elements of P-fibres to token-elements of the target. Both maps have to be compatible with the incidence morphisms. The nets from from Figure 1 will serve as our running example in Chapter 3 - 5.

## 3  The two Petri topologies of a net

Graph-theoretical properties of an undirected net can be characterized by topological methods, too. This insight marked already the early period of Petri net theory ([Fer1975], [GLT1980]).

### 3.1    Definition (*Petri space*)

A topological space X is a *Petri space*, iff it satisfies the following two conditions:
- Arbitrary intersections of open subsets are open, i.e. for any – not necessarily finite - family $(U_i)_{i \in I}$ of open sets $U_i$ the intersection $\bigcap_{i \in I} U_i$ is open again.
- Every point of X is either open or closed.

The concept of a Petri space expresses the fundamental duality of the bipartition as a topological duality.

### 3.2    Proposition (*The two Petri topologies of a net*)

For an undirected net N = ( X, ad ) the adjacency relation ad $\subset$ X x X defines two Petri topologies on the set X of nodes:

- A subset U $\subset$ X is open with respect to the P-*topology*, iff
$$[ p \in X \text{ and } (p, t) \in ad \text{ for } t \in U ] \Rightarrow p \in U.$$
- A subset A $\subset$ X is open with respect to the T-*topology*, iff
$$[ t \in X \text{ and } (p, t) \in ad \text{ for } p \in A ] \Rightarrow t \in A.$$

Open sets of the P-topology are the place-bordered subnets, while transition-bordered subnets are open with respect to the T-topology. If a set is open with respect to one of both topologies, then it is closed with respect to the other – and vice versa. Topological statements in this paper will always refer to the P-topology. We denote by $j_P: P_X \to X$ the embedding with respect to the P-topology of the discrete subspace of open points, called *places*. Analogously we denote by $j_T: T_X \to X$ the embedding with respect to the T-topology of the discrete subspace of closed points, called *transitions*. Correspondingly, for a subset $Q \subset X$ we set $T_Q := Q \cap T_X$ und $P_Q := Q \cap P_X$.

### 3.3    Proposition (*Canonical basis of a Petri space*)

For a Petri space X the family of open sets $( p )_{p \in P_X} \cup ( \tilde{t} )_{t \in T_X}$ is a basis of the P-topology, its *canonical basis*. Here $\tilde{t} := \{ t \} \cup \text{pre}(t) \cup \text{post}(t)$ denotes the smallest neighbourhoud of the closed point t with respect to the P-topology. The *canonical basis* of the T-topology has as members either a single transition or the set $\bar{p} = \{ p \} \cup \text{pre}(p) \cup \text{post}(p)$, a place p together with its pre- and postset.

### 3.4    Definition (*Fibres of a morphism*)

Consider a continuous map between two Petri spaces f: X $\to$ Y.

i) The inverse image of a point y ∈ Y, considered as a subspace of X, is called the *fibre* of f over y and is denoted by $X_y := f^{-1}(y)$. It is called a *T-fibre*, iff y ∈ $T_Y$ is a transition, and a *P-fibre*, iff y ∈ $P_Y$ is a place. More generally, we denote by $X_Q := f^{-1}(Q)$ the inverse image of an arbitrary subset Q ⊂ Y.

ii) The map f is called *discrete*, iff all its fibres – equipped with the subspace topology inherited from X - carry the discrete topology.

A continous map between two Petri spaces f: X → Y is discrete, iff it respects the sorts, i.e. $f(P_X) \subset P_Y$ and $f(T_X) \subset T_Y$. Some authors call such maps a *folding*.

### 3.5 Running example *(Open resp. closed sets)*

Typical subsets of $N_X$, which are open with respect to the P-topology, are places and the P-fibre $N_{X,u}$. Typical closed subsets are transitions and the T-fibre $N_{X,a}$. The map f: X → Y is continous, but not discrete.

## 4 Coloured nets as sheaf-cosheaf pairs

> *Sheaf theory is the subject, in which you do*
> *topology horizontally and algebra vertically.*
> attributed to *M. Auslander*

A coloured net attaches a set of binding-elements to every transition and a set of token-elements to each of its places. Variation of transitions and places as a parameter establishes the common families of colours. These two families are a special case of the mathematical concept of a cosheaf (resp. a sheaf), which considers all closed (resp. all open) subnets as parameters.

First we define a presheaf and a precosheaf making use of the elegant language of category theory (Definition 4.1). Then we comment on these concepts – more down to earth – using a notation with families. Presheaves and precosheaves are only a transition stage, the final objects are sheaves and cosheaves. As a general reference we recommend ([Bre1997], [MM1994]). We denote by Ab the category of Abelian groups or **Z**-modules. It is an Abelian category, and the concept of an exact sequence is well-defined.

### 4.1 Definition (*Presheaf and precosheaf*)

Consider a fixed topological space X as a category P(X): Objects are the open subsets of X. For morphisms between two open subsets there exist two possibilities: If one set contains the other, then there is one single morphism, the embedding. Otherwise the set of morphisms is empty.

- A *presheaf* (resp. *precosheaf*) of Abelian groups on X is a contravariant (resp. covariant) functor

$$\mathcal{F}: P(X) \to Ab.$$

- A *morphism* between two presheaves (resp. precosheaves) is a natural transformation between the two functors.

The essential property of the functor is a set of restrictions (resp. extensions) between the Abelian groups belonging to different open subsets. These morphisms have to respect composition and identities.

### 4.2 Remark (*Presheaf and precosheaf*)

i) A presheaf $\mathcal{C}$ of Abelian groups on a topological space X is a family of Abelian groups $(\mathcal{C}(U))_{U \subset X \text{ open}}$ and a family of morphisms in Ab, called *restrictions*, $r_{V,U}: \mathcal{C}(U) \to \mathcal{C}(V)$, V ⊂ U ⊂ X open, with

$$r_{U,U} = id \text{ and } r_{W,V} \circ r_{V,U} = r_{W,V} \text{ for } W \subset V \subset U \subset X \text{ open}.$$

ii) A precosheaf $\mathcal{B}$ of Abelian groups on X is a family of Abelian groups $(\mathcal{B}(U))_{U \subset X \text{ open}}$ and a family of morphisms in <u>Ab</u>, called *extensions*, $e_{U,V}: \mathcal{B}(V) \to \mathcal{B}(U)$, $V \subset U \subset X$ open, with

$$e_{U,U} = \text{id and } e_{U,V} \circ e_{V,W} = e_{U,W} \text{ for } W \subset V \subset U \subset X \text{ open.}$$

iii) The elements of the Abelian groups $\mathcal{C}(U)$ (resp. $\mathcal{B}(U)$) are called the *sections* over U of the presheaf $\mathcal{C}$ (resp. the precosheaf $\mathcal{B}$). A common notation is $\Gamma(U, \mathcal{C}) := \mathcal{C}(U)$.

### 4.3 Remark *(Change of coefficients and positivity)*

Besides the category <u>Ab</u> we consider presheaves and precosheaves with values in different categories. They arise from extending the coefficients of the sections with respect to a distinguished basis along the chain of inclusions $N \subset Z \subset Q$. The algebraic operation for the extension $Z \subset Q$ is the tensor product. It extends a $Z$-module B to the $Q$-vector space $B_Q := B \otimes_Z Q$ and similar presheaves and precosheaves. The cone of *non-negative sections* is formed by sections with coefficients from $N$ resp. $Q^+$. A morphism, which maps non-negative sections to non-negative ones, will be called *signed morphism*.

A sheaf adds to the properties of a presheaf the possibility to glue global sections from local ones. A section of a sheaf is characterized by a product of local sections, which coincide after restriction on their common domain of definition. Analogously, a section of a cosheaf is represented by a sum of local sections. It is an equivalence class of a coproduct of local sections. The equivalence relation is generated by sections, which extend a common element on the intersections.

### 4.4 Definition *(Sheaves and cosheaves)*

Denote by X a topological space.

i) A presheaf $\mathcal{C}$ of Abelian groups on X is a *sheaf*, iff for every open set $U \subset X$ and every open covering $(U_i)_{i \in I}$ of U the following sequence in <u>Ab</u> is exact

$$0 \to \mathcal{C}(U) \xrightarrow{r} \Pi_{i \in I} \mathcal{C}(U_i) \xrightarrow{s} \Pi_{j,k \in I} \mathcal{C}(U_j \cap U_k).$$

Both morphisms to a product are determined by the corresponding morphisms to the factors

$$r_i := r_{U_i, U}: \mathcal{C}(U) \to \mathcal{C}(U_i)$$

$$s_{jk}: \Pi_{i \in I} \mathcal{C}(U_i) \longrightarrow \mathcal{C}(U_j \cap U_k), \quad s_{jk}((f_i)_{i \in I}) := r_{U_j \cap U_k, U_j}(f_j) - r_{U_j \cap U_k, U_k}(f_k).$$

ii) A precosheaf $\mathcal{B}$ of Abelian groups on X is a *cosheaf*, iff for every open set $U \subset X$ and every open covering $(U_i)_{i \in I}$ of U the following sequence in <u>Ab</u> is exact

$$\coprod_{j,k \in I} \mathcal{B}(U_j \cap U_k) \xrightarrow{d} \coprod_{i \in I} \mathcal{B}(U_i) \xrightarrow{e} \mathcal{B}(U) \to 0.$$

Both morphisms from the coproduct are determined by corresponding morphisms from the summands

$$d_{jk}: \mathcal{B}(U_j \cap U_k) \longrightarrow \coprod_{i \in I} \mathcal{B}(U_i), \quad d_{jk} := e_{U_j, U_j \cap U_k} - e_{U_k, U_j \cap U_k}$$

$$e_i := e_{U, U_i}: \mathcal{B}(U_i) \to \mathcal{B}(U).$$

iii) A *morphism* between two sheaves (resp. cosheaves) is a morphism between the corresponding presheaves (resp. precosheaves).

### 4.5 Remark *(Sheaf and cosheaf as limit resp. colimit)*

For readers interested in the categorical background: Definition 4.4 expresses the characteristic glueing property of a sheaf (resp. cosheaf) as a limit (resp. colimit). Global sections $\mathcal{C}(U)$ resp. $\mathcal{B}(U)$ are the equalizer of $\left[\prod_{i \in I} \mathcal{C}(U_i) \rightrightarrows \prod_{i,j \in I} \mathcal{C}(U_i \cap U_j)\right]$ resp. the coequalizer of $\left[\coprod_{i,j \in I} \mathcal{B}(U_i \cap U_j) \rightrightarrows \coprod_{i \in I} \mathcal{B}(U_i)\right]$. Making use of these types of limits one can define sheaves and cosheaves with values in arbitrary categories C. If each object of C has an underlying set, then the forgetful functor transforms a presheaf with values in C into a presheaf with values in Set, the category of sets. Because the functor respects limits, it respects the sheaf property. The forgetful functor does not respect colimits. Therefore a cosheaf with values in C has no underlying cosheaf of sets in general. Instead one has to use the adjoint functor. E.g. the functor, which attaches to a set the corresponding free algebraic object from C (Abelian group, Abelian monoid, etc.), transforms a cosheaf of sets into a cosheaf with values in C ([Ber1991]). To determine a sheaf or a cosheaf, it is not necessary to consider in Remark 4.2 or Definition 4.4 all open sets. It suffices to fix the sections over the elements of a basis. For a Petri space we will always take the canonical basis from Proposition 3.3.

Using a continous map between topological spaces one can map (pre)sheaves and (pre)cosheaves in a covariant way from one topological space to the other. Sections of the direct image sheaf over an open set are sections of the original sheaf over the inverse image of this set. The inverse image of an open set under a continous map is by definition open again.

### 4.6     Proposition (*Direct image of sheaves and cosheaves*)

Consider a continous map f: X → Y between two topological spaces. Let $\mathcal{F}$ be a precosheaf and $\mathcal{G}$ a presheaf on X. The covariant functor $U \mapsto \mathcal{F}(X_U)$ on open sets $U \subset Y$, is a precosheaf $f_*\mathcal{F}$ on Y, the *direct image* of $\mathcal{F}$. If $\mathcal{F}$ is a cosheaf, then also $f_*\mathcal{F}$ is a cosheaf. The contravariant functor $U \mapsto \mathcal{G}(X_U)$, on open sets $U \subset Y$, is a presheaf $f_*\mathcal{G}$ on Y, the *direct image* of $\mathcal{G}$. If $\mathcal{G}$ is a sheaf, then also $f_*\mathcal{G}$ is a sheaf.

The direct image is a natural transformation between precosheaves: A morphism between precosheaves on X induces a morphism between the corresponding direct images on Y. This attachment is compatible with composition. An analogous statement holds for presheaves. For a Petri space one can define the concept of a morphism between a (pre)cosheaf with respect to the T-topology and a (pre)sheaf with respect to the P-topology.

### 4.7     Definition (*Morphism between a precosheaf and a presheaf*)

Denote by X a Petri space. A *morphism* w: $\mathcal{B} \to \mathcal{C}$ between a precosheaf $\mathcal{B}$ of Abelian groups on the closed sets of X and a presheaf $\mathcal{C}$ of Abelian groups on the open sets of X is a family of morphisms of Abelian groups

$$w_{U,A}: \mathcal{B}(A) \to \mathcal{C}(U), \quad U \subset X \text{ open}, A \subset X \text{ closed},$$

such that for each pair of inclusions $A_1 \subset A_2$ and $U_1 \subset U_2$ the following diagram commutes

$$\begin{array}{ccc} \mathcal{B}(A_2) & \xrightarrow{w_{U_2,A_2}} & \mathcal{C}(U_2) \\ e_{A_2,A_1} \uparrow & & \downarrow r_{U_1,U_2} \\ \mathcal{B}(A_1) & \xrightarrow{w_{U_1,A_1}} & \mathcal{C}(U_1). \end{array}$$

Here $e_{A_2,A_1}$ denotes the extension of $\mathcal{B}$ and $r_{U_1,U_2}$ the restriction of $\mathcal{C}$. A morphism between a cosheaf and a sheaf is a morphism between the corresponding precosheaf and presheaf.

In this paper we take the following definition of a coloured net as our starting point ([Jen1982]):

### 4.8 Definition (*Tuple-notation of a coloured net*)

A *coloured net* in tuple-notation $(T, P, (B(t))_{t \in T}, (C(p))_{p \in P}, (w^{-/+}(t, p))_{(t,p) \in T \times P})$ comprises:

- Two non-empty, finite disjoint sets T of *transitions* and P of *places*
- a family of non-empty, finite sets $(B(t))_{t \in T}$ (*binding-elements*)
- a family of non-empty, finite sets $(C(p))_{p \in P}$ (*token-elements*)
- and two families $w^- = (w^-(t, p))_{(t,p) \in T \times P}$, $w^+ = (w^+(t, p))_{(t,p) \in T \times P}$ of *N*-linear maps between the free *N*-modules („bags") $B(t)_N$ and $C(t)_N$, the *negative* resp. *positive incidence functions*,

$$w^-(t, p), w^+(t, p) \in \mathrm{Hom}_N(B(t)_N, C(p)_N).$$

Now we are ready to translate the tuple-definition of a coloured net

$$(T, P, (B(t))_{t \in T}, (C(p))_{p \in P}, (w^{-/+}(t, p))_{(t,p) \in T \times P})$$

into the language of sheaf theory. We consider the underlying undirected net $X := T \cup P$ as a Petri space. We attach to every subset $A \subset T$ of transitions the coproduct of binding-elements $\coprod_{t \in A} B(t)$. This attachment is a cosheaf of sets $\coprod_{t \in T} B(t)$ on the subspace T. Similarly, by attaching to every subset $U \subset P$ of places the product of token-elements $\prod_{p \in U} C(p)$ we obtain a sheaf of sets $\prod_{p \in P} C(p)$ on the subspace P. In order to link both functors by the incidence morphisms, we map them along the corresponding embedding into the common Petri space X and extend their values from sets to free Abelian monoids.

### 4.9 Definition *(Coloured net as sheaf-cosheaf pair)*

A coloured net in tuple-notation has the following representation as *sheaf-cosheaf pair* ( $X, \mathcal{B}, \mathcal{C}, w^{-/+}$ ):

- Transitions and places form the *Petri space* $X := T \cup P$,
- binding-elements form the *cosheaf of bindings* $\mathcal{B} := j_{T*}\left(\coprod_{t \in T_X} B(t)\right)$ on the closed subsets of X,
- token-elements form the *sheaf of tokens* $\mathcal{C} := j_{P*}\left(\prod_{p \in P_X} C(p)\right)$ on the open subsets of X
- and the incidence functions form the two *negative* resp. *positive incidence morphisms*
$$w^{-/+}: \mathcal{B}_N \to \mathcal{C}_N \text{ with } w^{-/+}_{U,A}: \mathcal{B}(A)_N \to \mathcal{C}(U)_N, U \subset X \text{ open}, A \subset X \text{ closed},$$
which are induced by the universal property of coproduct and product from the incidence functions
$$w^{-/+}(t, p): B(t) \to C(p)_N, (t, p) \in A \times U.$$

### 4.10 Running Example *(Sheaf-cosheaf pair)*

The net $N_Y$ has two non-empty closed subsets, the set $A_1 = \{ a \}$ and the whole set $A_2 = Y$ and two non-empty open subsets $U_1 = \{ u \}$ and $U_2 = Y$. Sections of the cosheaf $\mathcal{B}_Y$ are the two binding-elements of the transition: $\mathcal{B}_Y(A_1) = \mathcal{B}_Y(A_2) = B_Y(a) = \{b_1, b_2\}$. The only non-trivial section of the sheaf $\mathcal{C}_Y$ is the token-element: $C_Y(U_1) = \mathcal{C}_Y(U_2) = C_Y(u) = \{c\}$. The families, which constitute the two incidence morphisms $w_Y^{-/+}: (\mathcal{B}_Y)_N \longrightarrow (\mathcal{C}_Y)_N$, contain the morphisms $w_{Y;U_1,A_1}^{-/+} = (2 \quad 2): \mathcal{B}_Y(A_1)_N \longrightarrow \mathcal{C}_Y(U_1)_N$ between the free Abelian monoids of rank 2 and 1.

The *state equation* of a coloured net introduces the difference $w := w^+ - w^-$ of the incidence morphisms. To this end one has to embed the free Abelian monoids of sections into their Abelian group extension. This task is performed by extending coefficients from $N$ to $Z$. The resulting cosheaf $\mathcal{B}_Z$ has sections $\mathcal{B}_Z(A) = \coprod_{t \in T_A} B(t)_Z$ and the resulting sheaf $\mathcal{C}_Z$ has sections $\mathcal{C}_Z(U) = \prod_{p \in P_U} C(p)_Z$. The state equation links binding and token-elements. The kernel of w are local flows, which glue to a sheaf $\mathcal{F}$ of Abelian groups on the closed sets of X. The cokernel of w are classes of markings with respect to potential reachability. They glue to a cosheaf $\mathcal{M}$ of Abelian groups on the open subsets of X. It suffices to define $\mathcal{F}$ and $\mathcal{M}$ over the members of the canonical basis.

### 4.11 Definition (*Flows and marking classes*)

Consider a coloured net $N = (X, \mathcal{B}, \mathcal{C}, w^{-/+})$ with *incidence morphism* $w := w^+ - w^- : \mathcal{B}_Z \to \mathcal{C}_Z$.

i) The sheaf $\mathcal{F}$ of *flows* on the closed subsets of X is defined for elements of the canonical basis as

$$\mathcal{F}(A) := \begin{cases} \ker[\, w_{p,\bar{p}} : \mathcal{B}_Z(\bar{p}) \longrightarrow \mathcal{C}_Z(p)\,] & A = \bar{p},\, p \in P_X \\ \mathcal{B}_Z(t) & A = \{t\},\, t \in T_X \end{cases}$$

with the restriction $\mathcal{F}(\bar{p}) \to \mathcal{F}(t) = B(t)$ for $t \in \bar{p}$ induced from the inclusion $\coprod_{u \in \bar{p}} B(u) \subset \prod_{u \in \bar{p}} B(u)$ composed with the canonical projection from the product. Sections from $\mathcal{F}(A)$ are flows of the restriction of the subnet induced by A.

ii) The cosheaf $\mathcal{M}$ of *marking classes* on the open subsets of X is defined for elements of the canonical basis as

$$\mathcal{M}(U) := \begin{cases} \mathrm{coker}[\, w_{\tilde{t},t} : \mathcal{B}_Z(t) \longrightarrow \mathcal{C}_Z(\tilde{t})\,] & U = \tilde{t},\, t \in T_X \\ \mathcal{C}_Z(p) & U = \{p\},\, p \in P_X \end{cases}$$

with the extension $\mathcal{M}(p) = C(p) \to \mathcal{M}(\tilde{t})$ for $p \in \tilde{t}$ induced from the canonical injection into the coproduct composed with the inclusion $\coprod_{q \in \tilde{t}} C(q) \subset \prod_{q \in \tilde{t}} C(q)$. Sections from $\mathcal{M}(U)$ are marking classes of the subnet induced by U.

The sheaf of flows and the cosheaf of marking classes generalize token-elements and binding-elements. Sections of $\mathcal{F}$ over a closed singleton form the free Abelian group with basis the binding-elements of the transition $\mathcal{F}(t) = \mathcal{B}(t)_Z$. Sections of $\mathcal{M}$ over an open singleton form the free Abelian group with basis the token-elements of the place in question $\mathcal{M}(p) = \mathcal{C}(p)_Z$.

### 4.12 Proposition (*Flows and marking classes*)

Consider a coloured net $N = (X, \mathcal{B}, \mathcal{C}, w^{-/+})$ over a Petri space X. From Definition 4.11 derives the following form of flows and marking classes over general closed resp. open sets.

i) A flow $\tau \in \mathcal{F}(A)$ over an arbitrary closed set $A \subset X$ is a family of binding-elements $\tau = \prod_{a \in T_A} b_a \in \prod_{a \in T_A} \mathcal{B}_Z(a)$, which satisfies for every place $p \in P_A$ the flow condition $\sum_{t \in \bar{p}} w_{p,t}(b_t) = 0$.

ii) Every marking class $\mu \in \mathcal{M}(U)$ over an arbitrary open subset $U \subset X$ can be represented by a family of

token-elements $c = \coprod_{u \in P_U} c_u \in \coprod_{u \in P_U} \mathcal{C}_Z(u)$. Another family $c' = \coprod_{u \in P_U} c'_u \in \coprod_{u \in P_U} \mathcal{C}_Z(u)$ represents the same marking class, iff there exists a step $b \in \mathcal{B}_Z(U)$ with $c - c' = w_{P_U, T_U}(b)$.

iii) The incidence morphisms $w^{-/+} : \mathcal{B}_N \longrightarrow \mathcal{C}_N$ induce on the level of sections a morphism of Abelian groups $w_{U,A}^{-/+} : \mathcal{F}(A) \longrightarrow \mathcal{M}(U)$ for each closed subset $A \subset X$ and open subset $U \subset X$.

### 4.13   Definition (*Marking and step*)

An *integer-valued marking* of a coloured net $N = (X, \mathcal{B}, \mathcal{C}, w^{-/+})$ is a section $\mu \in \Gamma(P_X, \mathcal{M})$ of the cosheaf of marking classes over the open set $P_X$ of all places. An *integer-valued step* is a section $\tau \in \Gamma(T_X, \mathcal{F})$ of the sheaf of flows over the closed set $T_X$ of all transitions. The subcone of non-negative sections defines markings and steps with coefficients from *N*.

### 4.14   Remark (*Marking and marking class, step and flow*)

For a coloured net $N = (X, \mathcal{B}, \mathcal{C}, w^{-/+})$ one has a surjective morphism $e_{X,P_X} : \Gamma(P_X, \mathcal{M}) \longrightarrow \Gamma(X, \mathcal{M})$, which projects markings onto global marking classes. The map annihilates those markings, which result from the firing of a step. Similarly one has an injective morphism $r_{T_X, X} : \Gamma(X, \mathcal{F}) \longrightarrow \Gamma(T_X, \mathcal{F})$, which embeds global flows into steps.

### 4.15   Running Example (*Flows and marking classes*)

Over the closed subsets $A_1 = N_{X,a}$ resp. $A_2 = X$ of the net $N_X$ the sheaf of flows $\mathcal{F}_X$ has as sections the free Abelian groups $\mathcal{F}_X(A_1) = \text{span}_Z < \tau_1, \tau_2 >$ resp. $\mathcal{F}_X(A_2) = \text{span}_Z < \tau_1, \tau_2, \tau_3 >$ with $\tau_3 = t_5 + t_6$. These groups have resp. rank 2 and 3. The inclusion $A_1 \subset A_2$ induces the canonical restriction $\mathcal{F}_X(A_2) \to \mathcal{F}_X(A_1)$. Over the open subset $U_1 = N_{X,u}$ the cosheaf of marking classes $\mathcal{M}_X$ has as sections the free Abelian group

$$\mathcal{M}_X(U_1) = \left( \bigoplus_{i=3,4} Z \cdot p_i^* \right) / \text{span}_Z < (1,-1) >$$ and over $U_2 = X$ the free Abelian group

$$\mathcal{M}_X(U_2) = \left( \bigoplus_{i=1,\ldots,4} Z \cdot p_i^* \right) / \text{span}_Z < (-1,-1,1,1), (1,0,-1,0), (1,1,-1,-1), (0,1,0,-1), (0,0,-1,1), (0,0,1,-1) >.$$ Both have rank 1, the class $\left[ p_3^* \right]$ serves as a base in both cases. The inclusion $U_1 \subset U_2$ induces an extension of marking classes $\mathcal{M}_X(U_1) \to \mathcal{M}_X(U_2)$ by extending markings by the zero-marking.

### 4.16   Remark (*Sheaf-cosheaf pair*)

i) What is the benefit to represent even uncoloured p/t nets as sheaf-cosheaf pair?

In a p/t net $N$ every transition has only a single binding-element, every place has only a single token-element. Hence, neither $\mathcal{B}$ nor $\mathcal{C}$ add to the topological properties of the net any algebraic information. Quite different is the significance of $\mathcal{F}$ and $\mathcal{M}$. Global sections $\Gamma(N, \mathcal{F})$ of $\mathcal{F}$ are the T-flows of the net. But being a sheaf, $\mathcal{F}$ provides much more information: It attaches to every transition-bordered subnet its T-flows, and it compares two subnets - one containing the other - by their restriction of flows. The sheaf $\mathcal{F}$ emphasizes the importance of the T-flows of all transition-bordered subnets as a structural invariant of the net.

Which structural net properties does the cosheaf $\mathcal{M}$ express? Global sections $\Gamma(N, \mathcal{M})$ are equivalence classes of markings with respect to potential reachability. The cosheaf $\mathcal{M}$ attaches in a functorial way to every place-bordered subnet all of its marking classes. If one subnet contains the other, then both are compared by the continuation of markings. The cosheaf $\mathcal{M}$ emphasizes the importance of marking classes with respect to potential reachability of all place-bordered subnets as a structural invariant of the net.

The sheaf-cosheaf approach takes into account also P-flows. A linear functional on marking classes maps two markings to the same value, if one is potentially reachable from the other. Hence the linear functionals on the sections of $\mathcal{M}$ are P-flows: The dual of the cosheaf $\mathcal{M}$ are the P-flows of a net and of all its place-bordered subnets.

ii) The identity between P-flows of a net and the dual of $\mathcal{M}$ holds for arbitrary coloured nets. The first step in the study of duality is to define the dual of a coloured net. A first result: The linear functionals on the marking classes of a net are the flows of the dual net, the marking classes of the dual net are the linear functionals on the flows of the original net. Hence P-flows of a net are T-flows on its dual and vice versa.

There are two fundamental concepts in linear net theory: *Flow* – in the sense of T-flow – and *marking class*. They are formalized by the sheaf $\mathcal{F}$ on the closed subnets and the cosheaf $\mathcal{M}$ on the open subnets. Here from the concept of P-flows derives by duality.

## 5   Morphisms of coloured nets

A morphism between coloured nets is a morphisms between two sheaf-cosheaf pairs. Hence it has as first component a continous map between the two Petri spaces involved. We define over this map two other morphisms. One maps the sheaf of flows of both nets, the other maps the cosheaf of marking classes. Both maps are linked by the incidence morphisms of the nets.

### 5.1   Definition (*Morphism between coloured nets*)

A *morphism* $(f, f_F, f_M): N_X \to N_Y$ between two coloured nets $N_X = (X, \mathcal{B}_X, \mathcal{C}_X, w_X^{-/+})$ and $N_Y = (Y, \mathcal{B}_Y, \mathcal{C}_Y, w_Y^{-/+})$ is formed by:

- A continous map $f: X \to Y$ of the underlying Petri spaces,
- a signed morphism $f_F: f_*\mathcal{F}_X \to \mathcal{F}_Y$ between the sheaves of flows
- and a signed morphism $f_M: f_*\mathcal{M}_X \to \mathcal{M}_Y$ between the cosheaves of marking classes,

which render commutative the following two diagrams - for $w^-$ and $w^+$ - with $T := T_{f(X)}$ and $P := P_{f(X)}$

$$\begin{array}{ccc}
f_*\mathcal{F}_X(T) & \xrightarrow{f_{F;T}} & \mathcal{F}_Y(T) \\
(f_*w_X^{-/+})_{P,T} \downarrow & & \downarrow w_{Y;P,T}^{-/+} \\
f_*\mathcal{M}_X(P) & \xrightarrow{f_{M;P}} & \mathcal{M}_Y(P)
\end{array}$$

The morphism is *discrete*, iff the continous map f is discrete.

### 5.2   Proposition (*Composition of morphisms*)

The composition of two morphisms of coloured nets is a morphism. If both are discrete, then the composition is discrete, too.

**Proof**. The proposition is true, because the direct image of sheaves and cosheaves is a functor with respect to continous maps, QED.

Two important types of morphisms are *abstractions* and *embeddings*. T-fibres of an abstraction are transition-refinements, P-fibres are place-refinements. An abstraction maps every flow of the refined net onto a binding-element of the abstracted net. Analogously, every token-element of the abstracted net arises as the image of a marking class of the refined net. An embedding extends injectively every binding resp. token-element of the subnet to a binding resp. token-element of the ambient net. The subnet contains with two nodes from the ambient net also all incident arcs.

### 5.3 Definition (*Abstraction, embedding*)

A morphism of coloured nets $(f, f_F, f_M): N_X \to N_Y$ is named

- *abstraction*, iff $f$, $f_F$ and $f_M$ are surjective,
- *embedding*, iff $f$ is an embedding of topological spaces, while $f_F$ and $f_M$ are injective.

### 5.4 Running Example (*Abstraction*)

With the notation of Example 4.10 and 4.15 we define the morphism $(f, f_F, f_M): N_X \to N_Y$ on the level of flows. The direct image has sections $(f_*\mathcal{F}_X)(a) = \mathcal{F}_X(A_1)$, $(f_*\mathcal{F}_X)(Y) = \mathcal{F}_X(A_2)$. We define

$$f_{F;a}: (f_*\mathcal{F}_X)(a) \longrightarrow \mathcal{F}_Y(a), \tau_i \mapsto b_i, i = 1, 2, \text{ and } f_{F;Y}: (f_*\mathcal{F}_X)(Y) \longrightarrow \mathcal{F}_Y(Y), \tau_i \mapsto \begin{cases} b_i & i = 1, 2 \\ 0 & i = 3 \end{cases}.$$ On the

level of marking classes we have $(f_*\mathcal{M}_X)(u) = \mathcal{M}_X(U_1)$, $(f_*\mathcal{M}_X)(Y) = \mathcal{M}_X(U_2)$. We define $f_{M;u}: (f_*\mathcal{M}_X)(u) \longrightarrow \mathcal{M}_Y(u), [p_3^*] \mapsto c$ and $f_{M;Y}: (f_*\mathcal{M}_X)(Y) \longrightarrow \mathcal{M}_Y(Y), [p_3^*] \mapsto c$. To prove the compatibility we have to verify the commutativity of

$$\begin{array}{ccc} f_*\mathcal{F}_X(a) & \xrightarrow{f_{F;a}} & \mathcal{F}_Y(a) \\ (f_*w_X^{-/+})_{u,a} \downarrow & & \downarrow w_{Y;u,a}^{-/+} \\ f_*\mathcal{M}_X(u) & \xrightarrow{f_{M;u}} & \mathcal{M}_Y(u) \end{array}$$

With respect to the bases introduced above this reduces to the matrix-equation $(2 \quad 2) \cdot \begin{pmatrix} 1 & 0 \\ 0 & 1 \end{pmatrix} = 1 \cdot (2 \quad 2)$.

All components of the morphism are surjective, hence it is an abstraction of coloured nets.

### 5.5 Remark (*Relation to the work of Lakos*)

Lakos identifies three forms of incremental changes for coloured nets. He formalizes each one by a certain type of morphism. In our interpretation they are special cases of our definition 5.3: A morphism, which *captures a subnet refinement* ([Lak1999], Def. 4.9), is an embedding with $f_F$ and $f_M$ being bijective. A morphism, which *captures a node refinement* ([Lak1999], Def. 4.7), is an abstraction; we consider ([Lak1999], Def. 3.12) to be a misprint. A morphism, which *captures a type-refinement* ([Lak1999], Def. 4.1), is an embedding with $f$ being the identity.

So far we have dealt with the structural aspect of coloured nets. A morphism of coloured nets maps the structural properties of a net, expressed by its topology, its flows and its marking classes. We now add a distinguished initial marking and consider the behavioural aspect of Petri nets. We require that a morphism of coloured Petri nets compares also the initial markings. As a consequence it maps even the behaviour, if there are no topological obstructions. These obstructions are transitions in the image with a pre- and postset, which is not fully contained in the image. Such transitions cannot appear, if the map is surjective or - at least - has open image.

In this paper we define the behaviour of Petri nets by occurrence sequences. A morphism maps *saturated* occurrence sequences and the change of marking, which results from their firing. Saturated occurrence sequences flip between two types of components: Occurrence sequences of one type occur within a T-fibre and have a flow as Parikh-vector. Occurrence sequences of the other type occur in a P-fibre. It is not excluded, that some components are zero. A saturated occurrence sequence does not change the marking within T-fibres, but it may change the marking across its border. This change is mapped to the image net.

### 5.6 Definition (*Saturated occurrence sequence*)

Denote by $(f, f_F, f_M): N_X \to N_Y$ a morphism of coloured nets. An occurrence sequence $\sigma$ of $N_X$ is *satu-*

*rated*, iff it is a catenation of finitely many occurrence sequences of $N_X$

$$\sigma = \sigma_{T,1} \sigma_{P,1} \ldots \sigma_{T,i} \sigma_{P,i} \ldots \sigma_{T,n} \sigma_{P,n},$$

with the following property: Every Parikh-vector $\tau(\sigma_{T,i}) \in \mathcal{F}_X(N_{a_i})$, $a_i \in T_Y$, is a flow. Every Parikh-vector $\tau(\sigma_{P,i}) \in \mathcal{B}_X(N_{u_i})$, $u_i \in P_Y$, belongs to a P-fibre. Due to the non-negativity of $f_F$, each Parikh-vector of a flow maps to a binding-element $f_{F;a_i}(\tau(\sigma_{T,i})) \in \mathcal{B}_Y(a_i)$, $i = 1,\ldots,n$, of $N_Y$ and the catenation of all these binding-elements defines an occurrence sequence, named $f_{F;T_Y}(\sigma)$.

### 5.7 Proposition *(Mapping of the behaviour)*

A morphism of coloured nets $(f, f_F, f_M): N_X \to N_Y$, which has open image $f(X) \subset Y$, maps an activated, saturated occurrence sequence of $N_X$ to an activated occurrence sequence of $N_Y$: If $\mu_{pre}$ is a marking and $\sigma$ a saturated occurrence sequence of $N_X$, then

$$\mu_{X,pre} \xrightarrow{\sigma} \mu_{X,post} \Rightarrow f_{M;P_Y}(\mu_{X,pre}) \xrightarrow{f_{F;T_Y}(\sigma)} f_{M;P_Y}(\mu_{X,post}).$$

**Proof.** It suffices to prove the proposition for occurrence sequences of length 1 within a T-fibre. Consider a transition $a \in T_Y$, a transition $t \in f^{-1}(a)$ and a binding-element $b \in \mathcal{B}_X(t)$, which is activated at the marking $\mu_{X,pre}$. For all places $p \in N_X$ holds $\mu_{X,pre}(p) = w^-_{N;p,t}(b) + \mu_+(p)$ with a token-element $\mu_+(p) \in \mathcal{C}_X(p)$. Because $f(X) \subset Y$ is open, also $\tilde{a} \subset f(X)$. For arbitrary, but fixed place $u \in \tilde{a}$ holds

$$f_{M;u}(\mu_{X,pre} \mid X_u) = f_{M;u}(w^-_{X;X_u,t}(b)) + f_{M;u}(\mu_+ \mid X_u) = w^-_{Y;u,a}(f_{F;a}(b)) + f_{M;u}(\mu_+ \mid X_u).$$

Variation of the place $u \in \tilde{a}$ as parameter shows $f_{M;P_Y}(\mu_{X,pre}) = w^-_{Y;P_Y,a}(f_{F;a}(b)) + f_{M;P_Y}(\mu_+)$. Hence the binding-element $f_{F;a}(b)$, which is the image of the original occurrence sequence, is activated at the marking $f_{M;P_Y}(\mu_{X,pre})$, which is the image of $\mu_{pre}$ under the signed map $f_M$. The final statement about the resulting marking follows from the state equation $\mu_{X,post} = \mu_{X,pre} + w_{X;P_X,t}(b)$ by applying $f_{M;P_Y}$:

$$f_{M;P_Y}(\mu_{X,post}) = f_{M;P_Y}(\mu_{X,pre}) + f_{M;P_Y}(w_{X;P_X,t}(b)) = f_{M;P_Y}(\mu_{X,pre}) + w_{Y;P_Y,a}(f_{F;a}(b)), \text{ QED}.$$

### 5.8 Definition (*Morphism of coloured Petri nets*)

A morphism of coloured Petri-nets $(f, f_F, f_M): (N_X, \mu_X) \to (N_Y, \mu_Y)$ is a morphism of the corresponding coloured nets $(f, f_F, f_M): N_X \to N_Y$, which satisfies the two following conditions:

- The composition with the extension map of the cosheaf

$$\Gamma(P_X, \mathcal{M}_X) \xrightarrow{e_{f^{-1}(P_Y),P_X}} \Gamma(f^{-1}(P_Y), \mathcal{M}_X) \xrightarrow{f_{M;P_Y}} \Gamma(P_Y, \mathcal{M}_Y)$$

maps the initial marking of $(N_X, \mu_X)$ to the initial marking of $(N_Y, \mu_Y)$: $f_{M;P_Y}(e_{f^{-1}(P_Y),P_X}(\mu_X)) = \mu_Y$

- and $f_{F;T_Y}$ maps every activated, saturated occurrence sequence $\sigma$ in $(N_X, \mu_X)$ to an activated occurrence sequence $f_{F;T_Y}(\sigma)$ of $(N_Y, \mu_Y)$: $[\mu_X > \sigma \Rightarrow [\mu_Y > f_{F;T_Y}(\sigma)$.

### 5.9 Remark (*Morphism of coloured Petri nets*)

i) The second requirement of Definition 5.8 is always satisfied, if the continous map f is surjective or has an open image (Proposition 5.7).

ii) A discrete morphism $(f, f_F, f_M): N_X \to N_Y$ satisfies $f^{-1}(P_Y) = P_X$ and $f^{-1}(T_Y) = T_X$. In this case the

extension map in Definition 5.8 is the identity and every occurrence sequence is saturated. The definition simplifies as follows: A discrete morphism defines a morphism $(f, f_F, f_M): (N_X, \mu_X) \to (N_Y, \mu_Y)$ of coloured Petri nets, iff $f_{M;P_Y}$ maps the initial markings and $f_{F;T_Y}$ maps activated occurrence sequences.

### 5.10 Definition (*Modification*)

A surjective, discrete morphism $(f, f_F, f_M): N_X \longrightarrow N_Y$ between two coloured net is a *modification*, iff both of the maps $f_F: f_*\mathcal{F}_X \to \mathcal{F}_Y$ and $f_M: f_*\mathcal{M}_X \to \mathcal{M}_Y$ are signed isomorphism. A modification will be called a *place-modification* (resp. *transition-modification*), iff all its T-fibres (resp. its P-fibres) are singletons.

A well known example is the unfolding of a coloured net by an uncoloured p/t net. The uncoloured net maps as a modification onto the coloured net. The map does not the change the behaviour of the net.

### 5.11 Proposition (*Invariance of the behaviour under modification*)

A modification $(f, f_F, f_M): N_X \longrightarrow N_Y$ between two coloured nets extends for arbitrary initial markings $\mu_X$ of $N_X$ to a morphism of Petri nets $(f, f_F, f_M): (N_X, \mu_X) \longrightarrow (N_Y, f_{M,P_Y}(\mu_X))$. This morphism induces a bijection of reachable markings.

**Proof**. A modification extends to a morphism of Petri nets by Proposition 5.7. It induces a bijective map of markings $\Gamma(P_Y, f_*\mathcal{M}_X) = \Gamma(f^{-1}(P_Y), \mathcal{M}_X) = \Gamma(P_X, \mathcal{M}_X) \xrightarrow{f_{M;P_Y}} \Gamma(P_Y, \mathcal{M}_Y)$, which respects the relation of reachability. Set $\mu_Y := f_{M;P_Y}(\mu_X)$ and consider an occurrence sequence $\mu_Y \xrightarrow{\sigma} \mu_{Y,post}$. Without restriction $\sigma$ is a single binding-element $b \in \mathcal{B}_Y(a)$ of a transition $a \in T_Y$. Because $f_{F;a}$ is a signed isomorphism, there is a unique non-negative flow $\tau \in \mathcal{F}_Y(N_a)$ with $f_{F;a}(\tau) = b$. It is activated at $\mu_X$. If $\mu_X \xrightarrow{\tau} \mu_{X,post}$, then $\mu_{Y,post} = f_{M;P_Y}(\mu_{X,post})$, QED.

### 5.12 Remark (*Relation to the work of Winskel*)

A *Winskel-morphism* $f = (\beta, \eta): N_X \to N_Y$ between two p/t nets in tupel-notation is a pair $(\beta, \eta)$: One component is a multirelation $\beta: P_X \to P_Y$ between the places, the other a partial function $\eta: T_X \to T_Y$ between the transitions and both satisfy $\beta(\text{pre}_X(t)) = \text{pre}_Y(\eta(t))$ and $\beta(\text{post}_X(t)) = \text{post}_Y(\eta(t))$ for all $t \in T_X$ ([WN1995]). The following differences between Winskel-morphisms and morphisms in the sense of Definition 5.1 attract notice:

- A Winskel-morphism is not necessarily globally defined. The domain of its multirelation and its partial function may be a proper, closed subset of the source net.
- A Winskel-morphism is not necessarily continous with respect to the Petri-topology: A given place of the source may be related to many different places of the target net.
- A Winskel-morphism has open codomain. Hence there are no obstructions against the mapping of the behaviour (Proposition 5.7).

It will turn out, that a Winskel-morphism can be represented by a morphism in the sense of Definition 5.1: The restriction of the Winskel-morphism on its domain and a subsequent modification of its codomain is a discrete morphism between coloured nets. The modification contracts those places of the codomain, which are related to the same place of the domain. The modification smoothes the discontinuities, but keeps the behaviour by colouring the contraction.

### 5.13 Proposition (*Morphism and Winskel-morphism*)

A Winskel-morphism $f = (\beta, \eta): N_{\tilde{X}} \longrightarrow N_{\tilde{Y}}$ has closed domain $X := \text{dom}(\beta) \cup \text{dom}(\eta) \subset \tilde{X}$ and open codomain $f(X) := \text{cod}(\beta) \cup \text{cod}(\eta) \subset \tilde{Y}$. There exists a place-modification $\pi: N_{\tilde{Y}} \longrightarrow N_Y$ onto a coloured net $N_Y$, such that the composition of relations $g := \pi \circ (\beta \cup \eta \,|\, X): X \longrightarrow Y$ is a continous function and extends to a discrete morphism of coloured nets $(g, g_F, g_M): N_X \longrightarrow N_Y$.

**Proof** (Sketch): ad i) From the definition $\beta(\text{post}_{\tilde{X}}(t)) = \text{post}_{\tilde{Y}}(\eta(t))$ and $\beta(\text{pre}_{\tilde{X}}(t)) = \text{pre}_{\tilde{Y}}(\eta(t))$ follow both statements, the closedness of the domain and the openess of the codomain, by direct verification.

ad ii) Denote by $N_X = (X, \mathcal{B}_X, \mathcal{C}_X, w_X^{-/+})$ the p/t net, which is generated by the closed subnet X of $N_{\tilde{X}}$. We define the coloured net $N_Y = (Y, \mathcal{B}_Y, \mathcal{C}_Y, w_Y^{-/+})$ as follows: Consider on $\tilde{Y}$ the equivalence relation, which is generated by the relation $(y_1 \sim y_2 :\Leftrightarrow \exists x \in P_{\tilde{X}} : y_1 \in \beta(x) \text{ and } y_2 \in \beta(x))$. The quotient topology on the set of equivalence classes $Y := (\tilde{Y} / \sim)$ is a Petri topology and the projection $\pi: \tilde{Y} \longrightarrow Y$ is continous, open and discrete. We define $\mathcal{B}_Y := \pi_* \mathcal{B}_{\tilde{Y}}$, $\mathcal{C}_Y := \pi_* \mathcal{C}_{\tilde{Y}}$ and $w_{Y;U,A}^{-/+} := \sum_{\pi(p) \in U} w_{\tilde{Y};p,A}^{-/+}$. By construction, the composition of the two relations $g := \pi \circ (\beta \cup \eta \,|\, X): X \longrightarrow Y$ is a map. Its continuity follows again by direct verification form the definition of a Winskel-morphism.

iii) By part i) and ii) the map $g: X \longrightarrow Y$ is continous and discrete. The closed Petri space $X \subset N_{\tilde{X}}$ generates a subnet $N_X$, which is a p/t net. The target net $N_Y$, which belongs to the Petri space Y, has 1-dimensional modules of bindings $\mathcal{B}_Y(a)_N \cong N$. But possibly the modules of token-elements $\mathcal{C}_Y(u)_N$ are higher-dimensional and the incidence morphisms $w_Y^{-/+}$ are represented by matrices. We define the morphism of sheaves $g_F: g_* \mathcal{B}_X \longrightarrow \mathcal{B}_Y$. We have $g_* \mathcal{B}_X(a) = \mathcal{B}_X(X_a) = \prod_{g(t)=a} B_X(t)_Z = \coprod_{g(t)=a} B_X(t)_Z$ for a closed singleton $a = g(t)$. The map $g_{F;a}: \coprod_{g(t)=a} B_X(t)_Z \longrightarrow B_Y(a)_Z$ is induced by the universal property of the coproduct from the identities $\text{id}: B_X(t)_Z \cong Z \longrightarrow B_Y(a)_Z \cong Z$. For a closed basic set $A = \overline{u}$ with a place $u \in N_Y$ holds $(g_* \mathcal{B}_X)(A) = \mathcal{B}_X(X_A) = \mathcal{B}_X\left(\bigcup_{g(p)=u} \overline{p}\right)$. We map a flow $\tau = \sum_i n_i t_i \in (g_* \mathcal{B}_X)(A)$, $n_i \in Z$ onto the element $g_{F;A}(\tau) = \sum_i n_i g(t_i) \in \mathcal{B}_Y(A)$. In order to show, that $g_{F;A}: (g_* \mathcal{B}_X)(A) \longrightarrow \mathcal{B}_Y(A)$ is well-defined, we have to prove, that $g_{F;A}(\tau)$ is a flow. The conditions on the multirelations of source and target of f imply by definition of the incidence morphism Y

$$\sum_i n_i \, \beta(\text{post}_X(t_i)) = \sum_i n_i \, \text{post}_{\tilde{Y}}(\eta(t_i)) = \sum_i n_i \, w_{Y;Y,t_i}^+(1) = w_{Y;Y,t_i}^+(g_{F;A}(\tau)).$$

Analogously $\sum_i n_i \, \beta(\text{pre}_X(t_i)) = w_{Y;Y,t_i}^-(g_{F;A}(\tau))$. The flow-condition $\sum_i n_i \, \text{post}_X(t_i) = \sum_i n_i \, \text{pre}_X(t_i)$ implies $\sum_i n_i \, \beta(\text{post}_X(t_i)) = \sum_i n_i \, \beta(\text{pre}_X(t_i))$. Hence $w_{Y;Y,\overline{q}}^+(g_{F,A}(\tau)) = w_{Y;Y,\overline{q}}^-(g_{F,A}(\tau)) = 0$ and $w_{Y;Y,\overline{q}}(g_{F,A}(\tau)) = 0$. It is left to the reader, to give an analogous definition for the morphism of cosheaves $g_M: g_* \mathcal{M}_X \longrightarrow \mathcal{M}_Y$, QED.

### 5.14  **Notations** *(Categories of coloured nets)*

We denote by <u>Net</u> the category of all coloured nets and morphisms according to Definition 4.9 and 5.1. Sheaves, cosheaves and morphisms from this category are defined over the monoid *N* and its Abelian group extension *Z*. Sections have natural or integer numbers as coefficients. Due to the linear structure of the category <u>Net</u> one can easily extend the ring of coefficients (Remark 4.3) and consider e.g. nets with rational coefficients. We denote by <u>Net</u>$_Q$ the category of rational coloured nets and their morphisms. We will omit the subscript for sheaves and cosheaves, if the context makes clear the ring of definition. The subcategories with only discrete morphisms are denoted by <u>Net-dis</u> resp. <u>Net-dis$_Q$</u>.

A Petri net is a marked net. We denote by <u>Pet</u> the category of coloured Petri nets and their morphisms according to definition 5.8. Here from derives the category <u>Pet$_Q$</u> by ring extension from the ring $Z$ to the field $Q$. In close analogy to the unmarked case we define the categories <u>Pet-dis</u> resp. <u>Pet-dis$_Q$</u>.

As a relativation of the categories considered so far we introduce the comma categories <u>Net-dis(N)</u> and <u>Net-dis(N)$_Q$</u> of relative coloured nets. Objects of these categories are discrete morphisms to a fixed coloured net N as basis. Morphisms of these categories are discrete morphisms, which commute with the basis morphisms of their domain and codomain. Products in these categories are called fibre products.

## 6 Products of coloured nets

In the present chapter we show, that all of the categories <u>Net-dis$_Q$</u>, <u>Pet-dis$_Q$</u> and <u>Net-dis(N)$_Q$</u> have products. Categorical products help to build complex nets from simpler one or to simplify the study of a net by first checking, if there exist some factors. First we construct the Kronecker product of coloured nets. If we restrict to discrete morphisms, then the Kronecker product has the universal property of the categorical product. The Kronecker product of two undirected nets is a subset of the topological product of their nodes. Namely the subset of sort-respecting pairs, formed either by two places or by two transitions. The pairing of a place with a transition is not allowed as a node of the product net. Due to the restriction to sort-respecting pairs also the morphisms of the category are restricted to discrete morphisms.

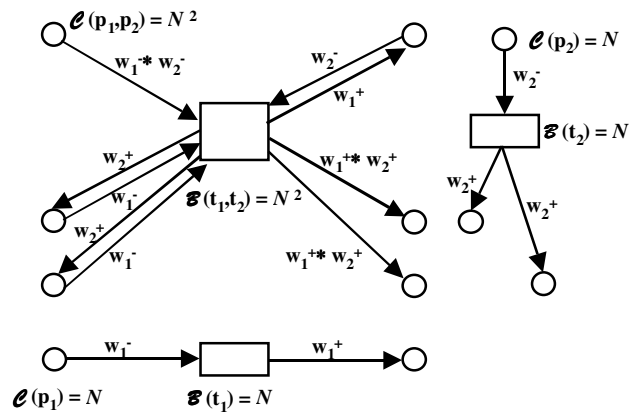

*Figure 2 A coloured net, which is the Kronecker product of two ordinary nets*

Products of nets and Petri-nets have been introduced by Dörfler in the setting of p/t nets and sort-respecting maps ([Dör1976]). They are named *Kronecker product*, because they are a close analogue to similar constructions from graph theory. Coloured nets had not yet been invented at the time of Dörflers paper. Our definition for the Kronecker product of coloured nets builds on Dörflers definition for the underlying undirected net and constructs on this base a coloured net. Figure 2 shows as an example the Kronecker product of two ordinary nets. The product is a coloured net, each node has a 2-dimensional module of colours. A transition of the product has two types of binding-elements: Either each binding-element of its factors separately or both as a synchronized mode. Both binding-elements keep their pre- and postsets from their factor.

### 6.1 Definition (*Kronecker product*)

The *Kronecker product* of two coloured nets $N_i = (X_i, \mathcal{B}_i, \mathcal{C}_i, w_i^{-/+})$, $i = 1, 2$, is the following coloured net

$$N := N_1 * N_2 := (X, \mathcal{B}_N, \mathcal{C}_N, w_N^{-/+}):$$

- The Petri space $X := X_1 * X_2 := (P_{X_1} \times P_{X_2}) \cup (T_{X_1} \times T_{X_2})$ carries the induced topology as a subspace $X \subset X_1 \times X_2$ of the product,

- sections of the cosheaf $\mathcal{B}_N$ of bindings over the product of closed sets are

$$\mathcal{B}_N(A_1 * A_2) := \coprod_{(a_1,a_2) \in T_{A_1} \times T_{A_2}} \left( \mathcal{B}_1(a_1) \prod \mathcal{B}_2(a_2) \right),$$

- sections of the sheaf $\mathcal{C}_N$ of tokens over the product of open sets are

$$\mathcal{C}_N(U_1 * U_2) := \prod_{(u_1,u_2) \in P_{U_1} \times P_{U_2}} \left( \mathcal{C}_1(u_1) \prod \mathcal{C}_2(u_2) \right),$$

- the extensions of the cosheaf $\mathcal{B}_N$ and restrictions of $\mathcal{C}_N$ are induced by the product of the corresponding maps of both factors

- and the two incidence morphisms $w_N^{-/+} : \mathcal{B}_N \longrightarrow \mathcal{C}_N$ are induced over product sets by the maps

$$w_{N;U_1*U_2,A_1*A_2}^{-/+} : \coprod_{(a_1,a_2)} \left( \mathcal{B}_1(a_1) \prod \mathcal{B}_2(a_2) \right) \longrightarrow \prod_{(u_1,u_2)} \left( \mathcal{C}_1(u_1) \prod \mathcal{C}_2(u_2) \right),$$

which are sum and product of the incidence maps of both factors $w_{i;u_i,a_i}^{-/+} : \mathcal{B}_i(a_i) \longrightarrow \mathcal{C}_i(u_i), i = 1, 2$.

We denote the two *projections* of the topological product by $p_i: X_1 * X_2 \to X_i$, $(x_1, x_2) \mapsto x_i$, $i = 1,2$.

### 6.2 Remark (*Flows and marking classes of the Kronecker product*)

Let $N = (X, \mathcal{B}_N, \mathcal{C}_N, w_N^{-/+}) := N_1 * N_2$ denote the Kronecker product of two coloured nets. Over the product of two closed sets $A = A_1 * A_2$ a section $\tau = \coprod_{a=(a_1,a_2)} (\tau_{a,1} \prod \tau_{a,2}) \in \mathcal{B}_N(A)$ is a flow $\tau \in \mathcal{F}_N(A)$, iff its trace $p_{i,F;A}(\tau) := \sum_{a \in T_A} \tau_{a,i} \in \mathcal{F}_i(A_i)$, $i = 1,2$, is a flow in both factor nets. Variation of $A_i$ as a parameter defines two sheaf morphisms, the *trace of flows*, $p_{i,F} : p_{i*}\mathcal{F}_N \longrightarrow \mathcal{F}_i$, $i = 1,2$. To define the rational *trace of marking classes* $p_{1,M} : p_{1*}\mathcal{M}_N \longrightarrow \mathcal{M}_1 \otimes_Z Q$ one has to take into account, that the product reproduces each place of the first factor with multiplicity equal to the number of places of the second factor. This amplification on the level of topology has to be corrected by a corresponding division on the level of marking classes. Therefore we extend the domain of coefficients from $Z$ to the field $Q$. Over the product $U = U_1 * X_2$ with open set $U_1$ a marking class $\pi = \coprod_u (\pi_{u,1} \prod \pi_{u,2}) \in \mathcal{M}_N(U_1 * X_2) = (p_{1*}\mathcal{M}_N)(U_1)$ maps to

$$p_{1,M;U_1*X_2}(\pi) := \frac{1}{\operatorname{card}(P_{X_2})} \sum_{u \in U_1 * X_2} \pi_{u,1} \in \mathcal{M}_1(U_1) \otimes_Z Q.$$ Variation of $U_1$ as a parameter defines the morphism $p_{1,M}$ between rational cosheaves. Similarly we define the morphism $p_{2,M} : p_{2*}\mathcal{M}_N \longrightarrow \mathcal{M}_2 \otimes_Z Q$.

### 6.3 Proposition (*Projections of the Kronecker product*)

Both projections of the Kronecker product $(p_i, p_{i,F}, p_{i,M}) : N_1 * N_2 \longrightarrow N_i$, $i = 1, 2$, are morphisms in the category Net-dis$_Q$.

### 6.4 Proposition (*Categorical product for discrete morphism*)

The Kronecker product with its two projections is the product in the category Net-dis$_Q$.

**Proof**. We show that the Kronecker product $(N_1 * N_2, p_1, p_2)$ of two rational coloured nets together with the two projections from Remark 6.2 has the universal property of the product. We consider a rational coloured net $N_Y = (Y, \mathcal{B}_Y, \mathcal{C}_Y, w_Y^{-/+})$ and two discrete morphisms $(f_i, f_{i,F}, f_{i,M}) : N_Y \longrightarrow N_i$, $i = 1,2$. Due to the universal property of the topological product the two continous maps induce a continous map into the product $f := f_1 * f_2 : Y \longrightarrow X_1 \times X_2, y \mapsto (f_1(y), f_2(y))$. It is discrete and its image is contained in $X = X_1 * X_2$, because both factor maps respect the sorts. Moreover it is uniquely determined by the condition $p_i \circ f = f_i, i = 1, 2$. In the next step we extend the continous map f to a morphism of coloured nets $(f, f_F, f_M): N_Y \longrightarrow N_1 * N_2$. First we define the morphism of rational sheaves $f_F : f_* \mathcal{F}_Y \longrightarrow \mathcal{F}_N$; we drop the subscript $Q$. For a given transition $a = (a_1, a_2) \in X_1 * X_2$ factorizes

$$(f_* \mathcal{F}_Y)(a) = \mathcal{F}_Y(Y_a) = \mathcal{F}_Y(Y_{a_1} \cap Y_{a_2}) = \prod_{f_1(t)=a_1 \text{ und } f_2(t)=a_2} B_Y(t)$$

as a product of binding-elements. The canonical embedding of the intersection into the product

$$\prod_{f_1(t)=a_1 \text{ und } f_2(t)=a_2} B_Y(t) \longrightarrow \left(\prod_{f_1(t)=a_1} B_Y(t)\right) \prod \left(\prod_{f_2(t)=a_2} B_Y(t)\right)$$

induces a map

$$\mathcal{F}_Y(Y_a) = \mathcal{F}_Y(Y_{a_1} \cap Y_{a_2}) \longrightarrow \mathcal{F}_Y(Y_{a_1}) \prod \mathcal{F}_Y(Y_{a_2}), \tau \mapsto \tau_1 \prod \tau_2.$$

We define by composition

$$f_{F;a} := \left[ \mathcal{F}_Y(Y_a) \longrightarrow \mathcal{F}_Y(Y_{a_1}) \prod \mathcal{F}_Y(Y_{a_2}) \xrightarrow{f_{1,F;a_1} \prod f_{2,F;a_2}} \mathcal{F}_1(a_1) \prod \mathcal{F}_2(a_2) = \mathcal{F}_N(a) \right].$$

Due to the finiteness of all nets there is no difference between the product and the coproduct of sections over closed sets $A \subset X_1 * X_2$. Because the fibres are discrete we have

$$\mathcal{F}_Y(Y_A) \subset \mathcal{B}_Y(Y_A) = \prod_{a \in A} \mathcal{F}_Y(Y_a) \text{ and } \prod_{a \in A} \mathcal{F}_N(a) = \mathcal{B}_N(A).$$

We define the composition

$$f_{F;A} := \left[ \mathcal{F}_Y(Y_A) \longrightarrow \prod_{a \in A} \mathcal{F}_Y(Y_a) \xrightarrow{\prod f_{F;a}} \prod_{a \in A} \mathcal{F}_N(a) = \mathcal{B}_N(A) \right].$$

It remains to show, that the image of this map is a flow

$$f_{F,A}(\tau) \in \mathcal{F}_N(A) \subset \mathcal{B}_N(A) \text{ for } \tau = \prod_{a \in A} \tau_a \in \mathcal{F}_Y(Y_A) \subset \prod_{a \in A} \mathcal{F}_Y(Y_a).$$

Consider an arbitrary place $p \in A^o$ from the open kernel of A. Because both maps $f_i$, $i = 1,2$, are morphisms, we have $w_{i;p_i,a_i} \circ f_{i,F;a_i} = f_{i,M;p_i} \circ f_{i*} w_{i;Y_{pi},Y_{ai}}$. We conclude

$$w_{N;p,A}(f_{F;A}(\tau)) = \sum_{a \in A} w_{N;p,a}(f_{F;a}(\tau_a)) = \sum_{a \in A} \left( w_{1;p_1,a_1}(f_{1,F;a_1}(\tau_{a,1})) \prod w_{2;p_2,a_2}(f_{2,F;a_2}(\tau_{a,2})) \right) =$$

$$\sum_{a \in A} \left( f_{1,M;p_1}\left( (f_{1*} w_{1;Y_{p1},Y_{a1}})(\tau_{a,1}) \right) \prod f_{2,M;p_2}\left( (f_{2*} w_{2;Y_{p2},Y_{a2}})(\tau_{a,2}) \right) \right) =$$

$$\left( f_{1,M;p_1} \prod f_{2,M;p_2} \right) \left( \sum_{a \in A} \left( f_{1*} w_{1;Y_{p1},Y_{a1}}(\tau_{a,1}) \prod f_{2*} w_{2;Y_{p2},Y_{a2}}(\tau_{a,2}) \right) \right) =$$

$$\left( f_{1,M;p_1} \prod f_{2,M;p_2} \right) \left( \sum_{a \in A} \left( f_{1*} w_{1;Y_{p1},Y_{a1}} \prod f_{2*} w_{2;Y_{p2},Y_{a2}} \right)(\tau_{a,1} \prod \tau_{a,2}) \right) = \left( f_{1,M;p_1} \prod f_{2,M;p_2} \right) \left( (f_* w_{N;Y_p,Y_A})(\tau) \right) = 0.$$

By construction all local maps $f_{F,A}$ are compatible with restrictions. Hence they define a sheaf morphism $f_F : f_* \mathcal{F}_Y \longrightarrow \mathcal{F}_N$. It is uniquely determined by the equations $p_{i*}(f_F) \circ p_{i,F} = f_{i,F}, i = 1,2$. We leave to the reader the analogous definition $f_M : f_* \mathcal{M}_Y \longrightarrow \mathcal{M}_N$ and the check, that $(f, f_F, f_M): N_Y \longrightarrow N_1 * N_2$ is a morphism, which completes a commutative diagram, QED.

A universal object like the product is determined only up to a canonical isomorphism. In Net-dis$_Q$ we obtain different products by variation of the scaling factor in the definition of the traces. Now we introduce initial markings and consider Petri nets. We will select a distinguished representation by fixing the scaling factor with a condition on the initial markings.

### *6.5*     **Definition** *(Kronecker product of coloured Petri nets)*

The *Kronecker product* of two coloured Petri nets $(N_i, \mu_i)$, $i = 1,2$, is the coloured Petri net $(N_1, \mu_1) * (N_2, \mu_2) := (N_1 * N_2, \mu_1 * \mu_2)$ with the product marking

$$\mu_1 * \mu_2 \in \Gamma(P_X, \mathscr{M}_{N_1 * N_2}), (\mu_1 * \mu_2)(p_1, p_2) := \mu_1(p_1) \prod \mu_2(p_2) \in \mathcal{C}_1(p_1) \prod \mathcal{C}_2(p_2).$$

A marking $\mu$ of the product is *saturated*, iff $\mu = p_{1,M}(\mu) * p_{2,M}(\mu)$ with $p_{i,M}$, $i = 1,2$, the trace morphisms.

### *6.6*     **Proposition** *(Categorical product of Petri nets)*

The Kronecker product of Petri nets with its two projections is the product in the category Pet-dis$_Q$.

Even if both coloured Petri nets belong to Pet-dis, their product does not satisfy the universal property in this category, because the trace maps are only defined over *Q*. In the following Proposition 6.7 we show, that a marking in the product is reachable, iff both of its factor markings are reachable in the corresponding factor nets. If these factors belong to Pet-dis, then also the initial marking and all reachable markings in the product as well as their traces have coefficients from *Z*.

### *6.7*     **Proposition** *(Reachability in the product of Petri nets)*

Reachable markings in the product of two Petri nets from Pet-dis$_Q$ correspond bijectively to pairs of reachable markings in both factor nets. For factor nets from Pet-dis this bijection restricts to a bijection of integer-valued markings.

**Proof**. Denote by $(N, \mu) := (N_1, \mu_1) * (N_2, \mu_2)$ the Kronecker product of two coloured Petri nets. Both projections $(N, \mu) \to (N_i, \mu_i)$, $i = 1,2$, are morphisms of Petri nets by Proposition 5.7. Hence the product

$p_{1,M;P_1} \prod p_{2,M;P_2} : \Gamma(P_N, \mathscr{M}_N) \longrightarrow \Gamma(P_1, \mathscr{M}_1) \prod \Gamma(P_2, \mathscr{M}_2)$ restricts to a map of reachable markings

$p_{1,M;P_1} \prod p_{2,M;P_2} : [\mu\rangle \longrightarrow [\mu_1\rangle \prod [\mu_2\rangle$. The initial marking $\mu$ is saturated. In order to prove the injectivity of the map, we show that saturated occurrence sequences transform saturated markings into saturated markings. Consider a saturated occurrence sequence $\sigma$ with Parikh-vector $\tau = \tau(\sigma)$. We assume – without loss of generality – that $\sigma$ is the binding-element $\tau = \tau_1 \prod \tau_2 \in B_N(t) = B_1(t_1) \prod B_2(t_2)$ of a single transition $t = (t_1, t_2) \in N$. Denote by $\mu_{pre}$ a saturated marking, which activates $\tau$, and denote by $\mu_{post}$ the marking, which results from the firing of $\tau$ at $\mu_{pre}$. We apply $p_{i,M}$, $i = 1,2$, to the state equation $\mu_{post} = \mu_{pre} + w_{N;N,a}(\tau)$ and obtain $p_{i,M}(\mu_{post}) = p_{i,M}(\mu_{pre}) + w_{i;N_i,a_i}(p_{i,T}(\tau_i))$. Spelling out the components of the state equation shows

$$\mu_{post}(p) = (\mu_{pre,1}(p_1) \prod \mu_{pre,2}(p_2)) + (w_{1;p_1,a_1}(\tau_1) \prod w_{2;p_2,a_2}(\tau_2)) =$$
$$= (\mu_{pre,1}(p_1) + w_{1;p_1,a_1}(\tau_1)) \prod (\mu_{pre,2}(p_2) + w_{2;p_2,a_2}(\tau_2))$$

Comparing the result under variation of the place $p \in N$ shows $\mu_{post} = p_{1,M}(\mu_{post}) * p_{2,M}(\mu_{post})$. Hence the marking $\mu_{post}$ is saturated. In order to show the surjectivity of the map under consideration we start with two occurrence sequences $\sigma_i$, i = 1,2 from the factors. Without loss of generality, each of them is a single binding-element $\tau_i \in \mathcal{B}_i(t_i)$. Reversing the direction of the above arguments shows, that in the product the occurrence sequence $\sigma$ with the single binding-element $\tau = \tau_1 \prod \tau_2 \in \mathcal{B}_N(t) = \mathcal{B}_1(t_1) \prod \mathcal{B}_2(t_2)$ is activated at the product marking $\mu = \mu_1 * \mu_2$, QED.

### 6.8 Remark *(Diagonal morphism)*

According to Proposition 6.4 each coloured net N has a *diagonal morphism* id*id: N → N * N with the following universal property: For a coloured net $N_Y$ and two morphisms $f_i : N_Y \longrightarrow N$, i = 1,2, in Net-dis$_\rho$, the product $f_1 * f_2 : N_Y \longrightarrow N * N$ factorizes over the diagonal as $N_Y \longrightarrow N \xrightarrow{id*id} N * N$, iff $f_1 = f_2$. We prove, that the diagonal morphism maps N isomorphic onto a subnet of the product, i.e. it embeds N into the product.

### 6.9 Proposition *(Existence of the diagonal)*

For a coloured net N the diagonal morphism induces an isomorphism $\delta: N \to \Delta_N$ onto a subnet $\Delta_N \subset N * N$ (the *diagonal*).

**Proof.** We define the diagonal of $N = (X, \mathcal{B}, \mathcal{C}, w^{-/+})$ as the coloured net $\Delta_N := (\Delta, \mathcal{B}_\Delta, \mathcal{C}_\Delta, w_\Delta^{-/+})$:

- The Petri space $\Delta := \{(x,x): x \in X\} \subset X * X$ has the subspace topology
- Binding-elements over closed product sets are
  $\mathcal{B}_\Delta((A * A) \cap \Delta) := \Delta_{\mathcal{B}(A)}$, the diagonal of the **Z**-module $\mathcal{B}(A) \prod \mathcal{B}(A)$,
- Token-elements over open product sets are
  $\mathcal{C}_\Delta((U * U) \cap \Delta) := \Delta_{\mathcal{C}(U)}$, the diagonal of the **Z**-module $\mathcal{C}(U) \prod \mathcal{C}(U)$,
- The extensions and restrictions of the cosheaf $\mathcal{B}_\Delta$ resp. the sheaf $\mathcal{C}_\Delta$ as well as the incidence morphisms $w_\Delta^{-/+}$ result from the corresponding morphisms of $\mathcal{B}_{N*N}$ resp. $\mathcal{C}_{N*N}$.

The diagonal has the flows $\mathcal{F}_\Delta(A) = \begin{cases} \Delta_{\mathcal{B}(a)} & A = \{t\}, \text{transition } t = (a,a) \in \Delta_X \\ \Delta_{\mathcal{F}(\bar{u})} & A = \bar{p} \cap \Delta, \text{place } p = (u,u) \in \Delta_X \end{cases}$

and the marking classes $\mathcal{M}_\Delta(U) = \begin{cases} \Delta_{\mathcal{C}(u)} & U = \{p\}, \text{place } p = (u,u) \in \Delta_X \\ \Delta_{\mathcal{M}(\tilde{a})} & U = \tilde{t} \cap \Delta, \text{transition } t = (a,a) \in \Delta_X \end{cases}$.

In order to consider the diagonal as a subnet $\Delta_N \subset N * N$, we have to extend the embedding of topological spaces j: $\Delta \to X * X$ to a morphism of coloured nets. The morphism of sheaves $j_F : \mathcal{F}_\Delta \longrightarrow \mathcal{F}_{N*N}$ is defined over a basic closed set A as the embedding of $\mathcal{F}_\Delta(A)$ into the **Z**-module

$$\mathcal{F}_{N*N}(A) = \begin{cases} \mathcal{B}(a) \prod \mathcal{B}(a) & A = \{t\}, \text{transition } t = (a,a) \in \Delta \\ \ker[\mathcal{F}_{N*N}(\bar{p}) \longrightarrow \mathcal{F}_{N*N}(\bar{p} - \Delta)] & A = \bar{p} \cap \Delta \text{ place } p = (u,u) \in \Delta \end{cases}.$$

A similar embedding of modules defines the morphism of cosheaves $j_M : \mathcal{M}_\Delta \longrightarrow \mathcal{M}_{N*N}$. It is left to the reader, to verify that $(j, j_F, j_M): \Delta_N \to N * N$ is an embedding of coloured nets and that the diagonal morphism id*id: N → N * N restricts to an isomorphism $\delta: N \to \Delta_N$, QED.

### 6.10 Proposition *(Inverse image of discrete morphisms)*

For a discrete morphism $f : N_X \longrightarrow N_Y$ of coloured nets and a subnet $j_Y : N_Z \longrightarrow N_Y$ there exists a subnet $j_X : f^{-1}(N_Z) \longrightarrow N_X$, the *inverse image*, such that the following diagram is Cartesian

$$\begin{array}{ccc} f^{-1}(N_Z) & \xrightarrow{j_X} & N_X \\ f_Z \downarrow & & \downarrow f \\ N_Z & \xrightarrow{j_Y} & N_Y \end{array}$$

**Proof**. We define the coloured net $f^{-1}(N_Z) := N_V = (V, \mathcal{B}_V, \mathcal{C}_V, w_V^{-/+})$ as follows: The set-theoretical inverse image $V := f^{-1}(Z) \subset X$ inherits the subspace topology from X. Binding-elements of transition $t \in T_V$ are defined as $\mathcal{B}_V(t) := f_{F;f(t)}^{-1}(j_{F;f(t)}(\mathcal{B}_Z(f(t)))) \subset \mathcal{B}_X(t)$. Sections over arbitrary closed sets are $\mathcal{B}_V(A) := \coprod_{t \in T_A} \mathcal{B}_V(t)$. Token-elements of a place $p \in P_V$ are $\mathcal{C}_V(p) := f_{M;f(p)}^{-1}(j_{M;f(p)}(\mathcal{B}_Z(f(p)))) \subset \mathcal{C}_X(p)$. Sections over arbitrary open sets are $\mathcal{C}_V(U) := \prod_{p \in P_U} \mathcal{C}_V(p)$. The incidence morphisms $w_V^{-/+} : \mathcal{B}_V \longrightarrow \mathcal{C}_V$ are defined as restrictions of $w_X^{-/+} : \mathcal{B}_X \longrightarrow \mathcal{C}_X$. Diagram chasing shows, that these definitions determine a subnet $N_V \subset N_X$, which completes the above diagram as a Cartesian diagram, QED.

The existence of fibre products follows by a standard construction from the existence of products, the diagonal and the inverse image. A posteriori the diagonal and the inverse image turn out as special cases of the fibre product.

### 6.11   Proposition *(Fibre product of coloured nets)*

The category <u>Net-dis(N)</u>$_Q$ has a product, it is a subnet of the product in <u>Net-dis</u>$_Q$.

**Proof.** For two objects $N_i \xrightarrow{g_i} N$, $i = 1, 2$, from <u>Net-dis(N)</u>$_Q$ we define the coloured net

$$N_1 *_N N_2 := (g_1 * g_2)^{-1} \Delta_N \subset N_1 * N_2$$

as inverse image of the diagonal under the morphism $N_1 * N_2 \xrightarrow{g_1 * g_2} N * N$ (Proposition 6.10). It is an object of <u>Net-dis(N)</u>$_Q$ with $(g_1 * g_2)|N_1 *_N N_2 \longrightarrow \Delta_N \cong N$ as basis morphism. The existence of a morphism $f = f_1 *_N f_2 : N_Y \longrightarrow N_1 *_N N_2$, as required for the universal property, follows from the universal property of $N_1 * N_2$ and the factorization of the induced morphism $f_1 * f_2$ over the diagonal, QED.

## 7   Conclusion and future work

The class of morphisms studied in the present paper is much more general than the class of sort-respecting morphisms. The only restriction for the underlying maps is to be continous, while sort-respecting morphisms have discrete fibres in addition.

- An important open question concerns the existence of products in the category <u>Net</u>$_Q$, which has arbitrary continous, but not necessarily discrete morphisms.

- An interesting approach to morphisms between coloured nets is due to Keller [Kel2000]. He works in a strict categorical setting and formalizes a coloured net as the morphism, which maps its unfolding onto its underlying uncoloured net. It would be useful to compare Kellers representation with the sheaf-cosheaf approach. Also the sheaf-functor can be represented by a map over the underlying topological space (espace étalé).

- Abstracting from the origin of the sheaf-cosheaf pair in Chapter 4 one can try to generalize the topological and algebraic situation: A topological space merges two disjoint, discrete subspaces, they are linked by two morphisms between a cosheaf on one of this spaces and a sheaf on the other.

- Another type of generalization deals with infinite nets, which allow also infinitely many colour-elements for a given node. From a mathematical point of view one performs the field extension $Q \subset R$. Then one introduces the structure of topological vector spaces on the sections of the sheaves and cosheaves in question. What we need for further formalization is a list of relevant examples – besides the universal unfolding of p/t nets.

- A characteristic property of Petri nets is their duality. The dual of a coloured net is the result of interchanging the P-topology with the T-topology followed by the transition from modules to linear functionals and from linear maps to their duals. We have seen a first application of duality at characterizing P-flows.